\documentstyle[twocolumn,prl,aps,epsf]{revtex}

\def\AmS{{\protect\the\textfont2
       A\kern-.1667em\lower.5ex\hbox{M}\kern-.125emS}}

\makeatletter
\def\@pnumwidth{2em}
\makeatother
\begin{document}
\draft
\twocolumn[\hsize\textwidth\columnwidth\hsize\csname @twocolumnfalse\endcsname
\title{Dielectric Anomalies of Solid CO and N$_2$ in the Audio Frequency Range}
\author{S.\ Pilla, J.\ A.\ Hamida, K.\ A.\ Muttalib, and N.\ S.\ Sullivan}
\address{Department of Physics, University of Florida,
 Gainesville, FL 32611}
\date{\today}
\maketitle

\begin{abstract}
We report the first audio frequency dielectric constant measurements of CO
and N$_2$ in their
solid phases. We have observed several new features, including (1)
strong hysteresis effects 
above an onset temperature T$_{h} \simeq$ 42 K in the $\beta$-phase of
pure N$_2$, (2) absence of the 
expected short range antiferroelectric dipolar ordering in CO down to 4.2 K, 
and (3) anomalous temperature dependence of 
the dielectric constant in the $\alpha$-phases of both CO and N$_2$.
Quantum mechanical treatment 
of the molecular rotation explains some of the observed anomalies in the
$\alpha$-phase, but the strong hysteresis indicates that pure geometrical 
frustration plays a significant role in the $\beta$-phase.  
\end{abstract}

]
\narrowtext

\paragraph*{}

\begin{figure}[htbp]
\protect \centerline {\epsfxsize=3.3in \epsfbox{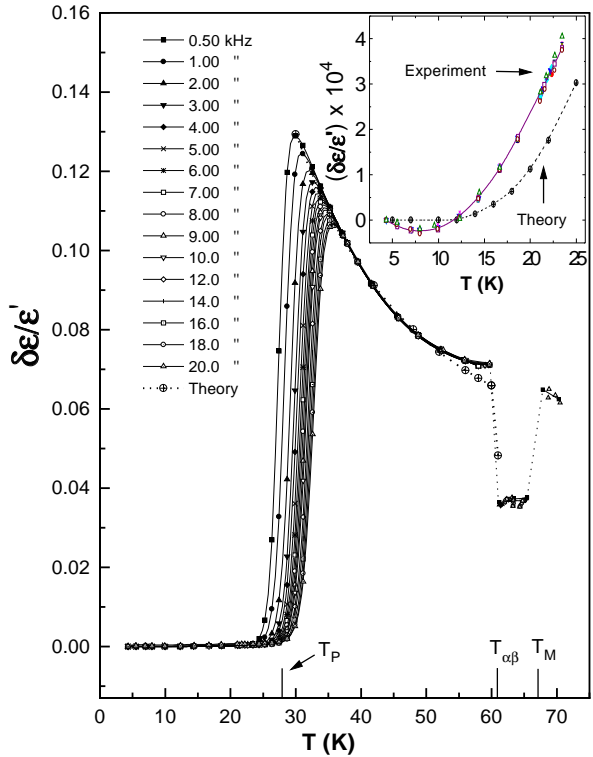}}
\caption{Dielectric constant of pure CO in the  audio frequency range relative to $\varepsilon'$, 
the value at 4.2 K. $\varepsilon' = 1.4211$ at 0.5 kHz. The inset shows the experimental and the 
theoretical results for the $\alpha$ phase below the dipolar freezing temperature $T_P$ 
in greater detail.}
\label{fig1}
\end{figure}

\begin{figure}[htbp]
\protect \centerline {\epsfxsize=3.3in \epsfbox{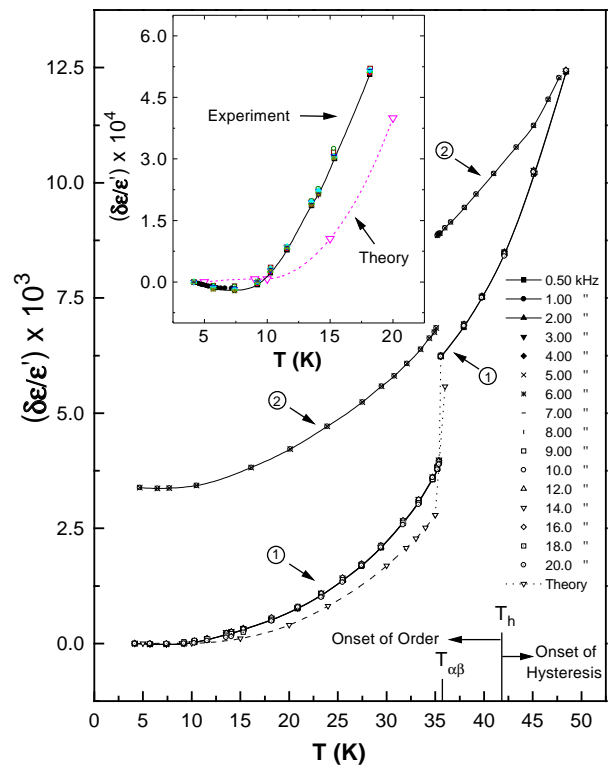}}
\caption{Dielectric constant of pure $^{14}$N$_2$ in the audio frequency range relative to $\varepsilon'$, 
the value at 4.2 K on curve (1). $\varepsilon' = 1.4255$ at 0.5 kHz. Curve (2) corresponds to 
$\delta\varepsilon/\varepsilon'$ of the same sample heated above $T_h$ and cooled.}
\label{fig2}
\end{figure}

Audio frequency measurements of $both$ the real and imaginary components
of the dielectric
susceptibility of solid CO and N$_2$ have been carried out for the first
time at temperatures 
ranging from their melting temperatures down to 4.2 K.
These simple molecular solids have been a subject of intensive
experimental and theoretical studies
during the last three decades (\protect
\cite{Scott,Hamida,Nicholls,Nary,Liu,koloskova,Manzhelii} 
and the references therein). The interest in these
systems derive from the fact that they serve as prototypes for studying the
underlying physics of the simplest glass formers. The interactions in
these solids are highly frustrated because the symmetries of
the interactions and that of the lattice geometry are 
incompatible and as a result new phenomena occur. 
The molecules interact via electrostatic quadrupole
moments which lead to a complex Pa3
anti-ferro-orientational ordering at low temperatures. 
The long range
ordering is however fragile, and with the introduction of 
relatively small
amounts of disorder (e.g.\ by replacing some N$_2$ molecules with non
interacting spherical Ar atoms) 
long range order is lost, and quadrupolar glass
states with randomly frozen orientations are 
observed. These systems are believed to be analogous to the spin glasses
and are particularly important because the interactions are short range
and well known, and the order parameters and molecular dynamics can be
studied directly.
\paragraph*{}
Dielectric constant/loss measurements in these systems are
particularly important because they directly probe the orientational
ordering. However, only measurements at microwave frequencies have been 
performed in the case of solid N$_2$ \protect \cite{Kempinski}. The molecular 
rotations
in these solids are strongly hindered, and as a consequence the relevant
time scales for collective molecular reorientation are expected to be 
much larger than those for an ideal free rotor ($\sim$10$^{-12}$ sec). 
In fact, glassy behaviors have been observed for solid N$_2$-Ar mixtures 
for which the characteristic reorientational time scale is $\sim$10$^{-4}$
sec \protect \cite{Hamida}. Measurements in this more interesting audio 
frequency 
range have not been performed due to the lack of required sensitivity. 
Nary et al. have carried out dielectric loss measurements of solid CO \protect 
\cite{Nary} 
in the audio frequency range, but to fully ascertain the nature of the dipolar
reorientations in the geometrically frustrated $\beta$ as well as the
ordered $\alpha$-phases, one also needs to study the dielectric constant 
in the entire temperature range of the solid phase. Moreover, CO has an
intrinsic dipole moment; while this makes it more accessible for
dielectric studies because of the large change in its dielectric
loss with temperature \protect \cite{Nary,Liu}, it also makes it more difficult 
to isolate
the much smaller contribution from the quadrupolar interactions which 
are responsible for the orientational ordering. We have developed a three 
terminal AC capacitance bridge \protect \cite{Adams,Anderson} with
two parts per billion sensitivity for measuring the real part of the
dielectric constant \protect \cite{Pilla} which allows us to study 
the collective orientational properties of solid quadrupolar molecular 
systems in the relevant audio frequency range.
\paragraph*{}
Figs.\ \ref{fig1} and \ref{fig2} show the real part of the dielectric constant 
$\varepsilon$ of CO and N$_2$ as a function of temperature $T$, in the 0.5 kHz 
to 20 kHz frequency range. The sensitivity and reliability of the apparatus can 
be judged from the observed jumps in $\varepsilon(T)$ at the structural phase 
transition at $T_{\alpha\beta}$, which is 61.57 K for CO and 35.61 K for N$_2$. 
At this temperature the low $T$ fcc phase, the orientationally ordered Pa3 
structure 
(or $\alpha$ phase), undergoes a first order transition to a high $T$ hcp or 
$\beta$ phase which does not support long range orientational order at any 
finite temperature. A second jump in $\varepsilon(T)$ occurs at the melting 
transition at $T_m$, which is at 68.13 K for CO and 63.15 K for N$_2$. Note 
that while the figures show $\varepsilon$ up to three 
decimal place accuracy, our apparatus has a much 
higher sensitivity, so the error bars for $\varepsilon$ 
are  negligible. However, as shown by the values of 
$T_{\alpha\beta}$ and $T_m$, the accuracy in the 
absolute value of temperature is within 0.3 K. 

In addition to the expected jumps, we observed several 
surprising features: 

1) For $T <$ 25 K, $\varepsilon(T)$ for polar CO and non-polar 
N$_2$ are similar, including the dip in $\varepsilon(T)$ near 7.5 K. This shows 
the absence of any dipolar 
ordering in CO down to 4.2 K. Also $\varepsilon(T)$ for CO did not depend on 
time (of the order of 
several days) which would otherwise indicate a slow thermal relaxation towards 
a dipolar ordered state. 
On the other hand the dielectric response differs markedly from the 
conventional temperature dependence 
for a non-polar material described by the Clausius-Mossotti (C-M) equation. 

2) For $T >$ 25 K and in the $\alpha$ phase, 
$\varepsilon(T)$ for CO is strikingly different from 
that of N$_2$. The difference is due primarily to the 
dipolar contribution from CO. The sharp rise at $T_P$ 
for CO can be attributed to the dipolar melting at $T_P$, above 
which dipoles begin to flip. We note that the results 
reported for  N$_2$-Ar-CO in \protect \cite{Liu} are
qualitatively indistinguishable from our data for pure CO in this
temperature range, except for a shift in
$T_P$, where CO was used as a tracer to increase the dielectric response. 
This shows the importance of probing molecular  
quadrupolar systems without a dipolar tracer. 

3) The most surprising feature for N$_2$ occurs in the 
$\beta$ phase, where it has a highly anomalous 
$\varepsilon(T)$. For thermal cycling below a temperature 
$T_h\approx$ 42 K, there is no hysteresis. In this regime, 
$\varepsilon(T)$ retraces a unique curve depending on the initial 
conditions, including 
the jump at $T_{\alpha \beta}$. However, if the system 
is taken above $T_h$ during a heating cycle (with the AC electric 
field present), the observed  $\varepsilon(T)$ follows a different 
curve during cooling, for which $\varepsilon(T)$ is higher than the 
previous curve. 
The difference between the two curves depends on 
how far above $T_h$ the system was heated. Fig.\ \ref{fig2} shows a typical 
hysteresis curve 
for a 1 kHz electric field of strength 1 kV/m where the system was 
heated to $\sim$48 K along curve (1) and cooled along curve (2). When the 
system is heated to 
a temperature less than 48 K but above $T_h$ and cooled, the $\varepsilon(T)$ 
lies between curves 
(1) and (2). Once the system is cooled below $T_h$, the corresponding 
$\varepsilon(T)$ curve remains reversible on thermal cycling, as long as the 
highest temperature 
remains below $T_h$. Within 0.1 K accuracy, $T_{\alpha\beta}$ remains the same 
for all curves. 
We should add that from these studies we cannot specify if $T_h$ is a 
sharp temperature associated with the onset 
of hysteresis or a crossover region. We also note that 
the thermal resistivity data on pure N$_2$ \protect \cite{koloskova} also has 
an anomaly close to T$_h$.

4) If the system is left isolated (in the absence of an AC electric field) at a 
temperature above $T_h$ for a 
long time (of the order of several 
hours), $\varepsilon(T)$ retraces the lowest curve, independent of 
the initial conditions.   

5) No hysteresis was observed for CO in the entire 
$\alpha$ phase. The small window of temperature 
between $T_{\alpha\beta}$ and $T_m$ prevented us from 
carrying out similar measurements in the $\beta$ phase of CO.  

6) The dielectric loss data for CO are similar to the 
results published
elsewhere \protect \cite{Nary,Liu} but the loss data for 
N$_2$ are featureless
for all frequencies and temperatures studied (not 
presented here).

7) The  frequency dependence 
of $\varepsilon$ at low temperatures 
can be collapsed on a unique curve when scaled by its 
value at 4.2 K. This is true for curves (1) and (2) and all 
other intermediate curves in Fig.\ \ref{fig2}. In general at 4.2 K, 
$\varepsilon$ increased with increasing 
frequency for both CO and N$_2$ ($\simeq $0.06\% increase at 20 kHz from its 
value at 0.5 kHz), 
but we do not understand this frequency dependence.  There is no frequency 
dependence for CO above $T_P$.
\paragraph*{}
While the $\beta$ phase of N$_2$ is novel, the behavior of $\varepsilon(T)$ in 
the
$\alpha$ phase is also unconventional. In order to understand the 
role of 
orientational ordering, we have developed a
phenomenological model
based on the known temperature dependence of the 
orientational order parameter for N$_2$ in 
the  $\alpha$ phase. We find that although the
molecules are large,  we need to consider the discrete quantum mechanical
orientational levels to explain the slow increase in 
$\varepsilon(T)$ in this regime. In contrast, the
classical treatment gives only a linear temperature 
dependence. In the present model, the change in $\varepsilon(T)$ at 
$T_{\alpha\beta}$ to be close to 
the experimental value requires 
50\% residual orientational ordering in the $\beta$ 
phase. This is consistent with our conjecture that the ordering begins at $T_h 
> T_{\alpha\beta}$. 
The same description holds also for CO up to the dipolar freezing temperature 
$T_P$. Above 
$T_P$ the dipolar contribution overwhelms the 
orientational contribution. The sharp 
rise at $T_P$ is associated with the onset of the flipping of the 
dipoles by
180$^{\circ}$. However, the 
molecules are not yet free to rotate and are
locked in the Pa3 structure. The subsequent 
decrease is therefore slower than the characteristic 
$1/T$ dependence expected for free dipoles. We find 
that a phenomenological model that accounts for the 
flipping without free rotation can describe the slower fall off qualitatively
and predict the jump at $T_P$. 
For CO, the drop in $\varepsilon$ at $T_{\alpha\beta}$ is consistent with the 
change in the order parameter 
with no residual ordering in the 
$\beta$ phase. Below we briefly sketch the 
phenomenological model for the $\alpha$ phases of CO and N$_2$.
\paragraph*{}
The average potential energy of a diatomic molecule at a site in a crystal with 
Pa3 ordering can be 
written as 
\begin{equation}
V=C\psi P_{2}(\cos\theta),
\end{equation}
in which $\theta$ is the angle between the molecular axis and the
crystal axis ($\hat{Z}$) at the lattice site and 
$P_{2}(\cos\theta)=-\frac{1}{2} +\frac{3}{2}\cos^{2}(\theta)$ \protect 
\cite{Kohin}. 
The quantity $\psi$ is the average value
of $P_{2} (\cos\theta_{j})$ for the neighboring molecules \it j. C \rm ($\simeq 
-673$ K for CO 
and $\simeq -520$ K for N$_{2}$) is a function of the molecular parameters and 
the lattice constants
\protect \cite{Kohin}. The order parameter $\psi$ is known experimentally for 
both CO and N$_{2}$ 
in the $\alpha$ phase but little is known about the residual ordering in the 
$\beta$ phase. If 
we write $\alpha = \alpha_{iso} + \Delta \alpha$ where $\alpha_{iso}$ is the 
isotropic
component and $\Delta \alpha$ is the anisotropic component of the
electric polarizability of a single molecule in the condensed phase, the
volume polarization can be written as \protect \cite{Wallace}:
\begin{equation}
{\bf P}= (1+ \frac{\chi_{e}}{3}) N 
(\alpha_{iso} 
+\Delta\alpha \langle\langle\cos^{2}
(\Omega -\theta)\rangle\rangle) {\bf E}_{ext}, 
\end{equation}
where $\chi _{e}$ is the electric susceptibility, $N$ is the number density, 
$\Omega$ is the angle
between the external electric field ($\hat{\bf E} _{ext}$) and $\hat{Z}$, 
and $\langle\langle\cdots\rangle\rangle$ refers to both configurational
and thermal average. In the present experiment at constant volume,  $N$
is a constant. Using the known values of the order parameter $\psi$
\protect \cite{Manzhelii}, one can calculate $\varepsilon
=1+\chi_e$.
However these calculated values of $\varepsilon$ increase linearly with
$T$ and do not agree with our experimental results. We note that when 
the separation
of the adjacent rotational energy levels of a rigid rotor are larger 
than $kT$, the classical Boltzman
distribution used in the averaging in (2) may not be appropriate. In fact
the separation of the two lowest energy levels ($\Delta E/k$) is $\simeq$
93 K for N$_2$ and $\simeq$ 104 K for CO (setting $\psi$ = 1) \protect 
\cite{Kohin}. 
From the Schr\"{o}dinger equation for a rigid rotor with moment of inertia $A$ 
and 
moving in the potential given by  Eq. 1, 
the $\theta$ dependence is given by 
\begin{equation}
\frac{d}{d\omega} \biggl[(1-\omega ^2 ) \frac{dQ}{d\omega}
\biggr]  \! + \! Q \biggl[ \mu - \! \lambda \omega ^2 \! - \! m^2 (1- \omega ^2 
)^{-1} \!
\biggr] \! = \!0,
\end{equation}
where $\omega = \cos\theta$, $\mu\hbar ^2 = A C \psi + 2AE$, 
$\lambda \hbar ^2 = 3A C \psi$, and $E$ is the energy. Eq. (3) is a spheroidal 
wave equation
and the eigensolutions Q$_{lm}(\lambda, \omega)$ are oblate spheroidal
wave functions. $\lambda$ is $\simeq -370$ K for CO and $\simeq -275$ K for 
N$_{2}$ 
\protect \cite{Kohin}. For such large values of $\lambda$, an expansion of the 
eigenfunctions 
in terms of Laguerre polynomials is appropriate. 
The eigenfunctions $Q_{lm}(\lambda, \omega)$ and the eigenvalues $\mu 
_{lm}(\lambda, \omega)$ for 
several of the lowest rotational states have been calculated using the method 
of 
Flammer\protect \cite{Flammer}. 
For various ($\psi$, $T$) sets, $\langle\langle\cos^{2}(\Omega 
-\theta)\rangle\rangle$ is calculated 
first by determining Q$_{lm}$ and $\mu _{lm}$ for several of the lowest 
eigenstates and the expectation
values of $(\cos^{2}(\Omega - \theta))_{lm}$ for each of these states. Then
\begin{equation}
\langle\langle\cos^{2}
(\Omega - \theta)\rangle\rangle=\frac{1}{Z}\sum _{j}
\gamma _{j}(\cos^{2}(\Omega - \theta))_{j}
e^{-E_{j}/kT},
\end{equation}
where Z is the 
partition function, $\gamma _{j}$ is the degeneracy of the $j$th state 
and
$E_{j} = \frac{\hbar ^2}{2A} \Bigl( \mu _{j} - \frac{AC \psi}{\hbar ^2}\Bigr)$. 
The $\varepsilon$ calculated for N$_2$ using Eqs.\ (2,4) for various 
temperatures, is shown as the dashed 
curve in Fig.\ \ref{fig2}. This is in better qualitative agreement with the 
experiment. In order to 
obtain the correct jump at $T_{\alpha \beta}$, we need to set $\psi$ = 0.5 in 
the $\beta$ phase. 
The same analysis has been carried out for CO with the known order 
parameter, and the resulting $\varepsilon(T)$ is shown as the dashed curve 
in Fig.\ \ref{fig1}(inset). Again the agreement with the experiment is 
qualitatively 
better. We mention two weaknesses of the above model. First, both N$_2$ 
and CO show a small dip in $\varepsilon(T)$ near 7.5 K which remains 
unaccounted for in the 
present model. Second, if one takes the $\alpha_{iso}$
for a free gaseous molecule \protect \cite{Maroulis,Gough}, and number density 
$N$ calculated from the 
lattice constants \protect \cite{Krupskii}, $\varepsilon'$ is about 15\%
and 11\% lower than the calculated value in the case of CO and $^{14}$N$_2$ 
respectively. 
In Figs.\ \ref{fig1} and \ref{fig2} we have
taken $\varepsilon'$ as the only adjustable unknown 
fitting parameter to compare the theory with the experiment. The lower 
experimental value 
compared to the theoretical estimate cannot be attributed to short-range 
antiferroelectric 
dipolar ordering in CO as suggested before \protect \cite{Nary} because N$_2$ 
also shows 
a similar behavior. However it can result from a lower polarizability of a 
molecule in 
the condensed phase than in the gaseous phase, or from a lower number density 
for a 
powdered sample than for a single crystal.
\paragraph*{}
In the case of CO above $T_P$, the dipolar contribution to the
dielectric constant should also be taken into consideration. Because the
molecules are locked in the Pa3
structure, the potential in the presence of an external electric field may be 
written as:
\begin{equation}
V= C \psi P_{2} (\cos\theta) - \mu _{\circ} |{\bf E}|
f(\psi) \cos(\Omega - \theta).
\end{equation}
The first part is the average potential energy of a single molecule as
in Eq. (1). ${\bf E}$ is the internal electric field, ${\bf
\mu_{\circ}}$ is the intrinsic dipole moment, and $f$ is some function
of the order parameter. Qualitatively, when $\psi$ is unity (i.e.\  the 
molecules are
completely ordered) the second term should be zero. When $\psi$ is zero
(i.e. the molecules are free to rotate) it should simply be $
\mu_{\circ}\cdot E$. For simplicity, we assume $f$ to be of the form
$(1-\psi)^{\tau}$, consistent with the above two limits. The volume 
polarization can be obtained for the above potential as usual. Taking the same 
values for the 
isotropic polarizability and number density which we used at 4.2 K, we find 
that for $f(\psi)=
(1-\psi)^{1/8}$ the experimental data can be fit very well up to
$T_{\alpha\beta}$ (the dotted curve in Fig.\ \ref{fig1}). We note that the 
predicted change in 
$\varepsilon(T)$ at $T_P$ is close to the change in the extrapolated value of 
the DC dielectric 
constant from experimental results within 0.8\%. We have no explanation for
the particular exponent in  $(1-\psi)^{1/8}$, but the qualitative
decrease which is slower than the characteristic $1/T$ dependence of free 
dipoles can be 
understood within the model. 
\paragraph*{}
In summary, we have made the first 
measurements of the dielectric constant of pure CO and N$_2$ in the
audio frequency range and observed several anomalous features. The low 
temperature behavior of both solids are very similar, but they are 
nevertheless very distinct from the conventional non-polar materials 
and can be qualitatively understood in terms of a phenomenological 
quantum mechanical model. The strong hysteresis effects in the $\beta$ 
phase of pure N$_2$ show that the conventional picture of the onset of 
ordering in these frustrated systems must be modified. 
Because the $\varepsilon(T)$ curve always retraces the lowest curve 
in Fig.\ \ref{fig2} whenever the
system is left isolated at a temperature above $T_h$ for a long time and then cooled, we
strongly believe that the above hysteresis effects cannot be explained 
by the lattice defects such as dislocations or vacancies. However, the results are 
consistent with a departure from ergodicity due to trapping in a limited 
region of configuration space. As a consequence of the frustration of 
the interactions, the configuration space has a rugged landscape with 
many quasi-equal low energy minima separated by potential energy 
barriers. This phenomenon is common to a wide class of glass formers 
and other frustrated systems with the introduction of disorder. The 
most significant result of this study is that the glassy effects are 
produced in a purely geometrically frustrated system without the 
introduction of disorder. This will be relevant to understanding the 
co-operative behavior in other geometrically frustrated systems 
(pyrochlores, spinels, and nuclear antiferromagnets).
\paragraph*{}
Stimulating discussions with Yasu Takano are acknowledged. We thank J.\ Xia,  
C. Parks, and P. Stachowiak for their help and gratefully acknowledge the 
support of all the personnel in our machine and electronics shops. This work 
is supported by a grant from the National Science Foundation No. DMR-962356.

\end{document}